\documentclass[pra,eqsecnum,amsmath,amsfonts,12pt,letterpaper,tightenlines,nofootinbib]{revtex4}

\usepackage{hyperref,graphicx}

\input{Qcircuit}

\newcommand{\braket}[2]{\langle #1 \vert #2 \rangle}
\newcommand{\abs}[1]{\lvert {#1} \rvert}
\def\ket#1{\lvert #1 \rangle}
\def\bra#1{{\langle #1 \rvert  }}
\def\tr{{\rm{tr}}}
\def\spec{{\rm{Spec}}}

\def\ba{\mathbf{a}}
\def\bb{\mathbf{b}}
\def\bc{\mathbf{c}}
\def\bx{\mathbf{x}}
\def\mz{\mathbb{Z}}

\def\sN{\mathcal{N}}
\def\sM{\mathcal{M}}

\newcommand{\be}{\begin{equation}}
\newcommand{\ee}{\end{equation}}
\newcommand{\ben}{\begin{eqnarray}}
\newcommand{\een}{\end{eqnarray}}
\newcommand{\bes}{\begin{subequations}}
\newcommand{\ees}{\end{subequations}}
\newcommand{\bF}{\begin{figure}}
\newcommand{\eF}{\end{figure}}

\def\ket#1{ | #1 \rangle}
\def\bra#1{{\langle #1 |  }}

\begin{document}

\title{Entanglement and the Power of One Qubit}
\author{Animesh Datta}
\email{animesh@unm.edu}
\author{Steven T. Flammia}
\email{sflammia@unm.edu}
\author{Carlton M. Caves}
\email{caves@info.phys.unm.edu}
\affiliation{Department of Physics and Astronomy, University of New Mexico,
    Albuquerque, NM 87131}
\date{\today}

\begin{abstract}
The ``Power of One Qubit'' refers to a computational model that has
access to only one pure bit of quantum information, along with $n$
qubits in the totally mixed state.  This model, though not as powerful
as a pure-state quantum computer, is capable of performing some
computational tasks exponentially faster than any known classical
algorithm.  One such task is to estimate with fixed accuracy the
normalized trace of a unitary operator that can be implemented
efficiently in a quantum circuit.  We show that circuits of this type
generally lead to entangled states, and we investigate the amount of
entanglement possible in such circuits, as measured by the
multiplicative negativity.  We show that the multiplicative negativity
is bounded by a constant, independent of $n$, for all bipartite
divisions of the $n+1$ qubits, and so becomes, when $n$ is large, a
vanishingly small fraction of the maximum possible multiplicative
negativity for roughly equal divisions.  This suggests that the global
nature of entanglement is a more important resource for quantum
computation than the magnitude of the entanglement.
\end{abstract}

\maketitle

\section{Introduction}
\label{S:intro}

Fully controllable and scalable quantum computers are likely many years
from realization.  This motivates study and development of somewhat
less ambitious quantum information processors, defined as devices that
fail to satisfy one or more of DiVincenzo's five criteria for a quantum
computer~\cite{DiVincenzo00}.   An example of such a quantum
information processor is a mixed-state quantum system, which fails to
pass DiVincenzo's second requirement, that the system be prepared in a
simple initial state.

The prime example of a mixed-state quantum information processor is
provided by liquid-state NMR experiments in quantum information
processing~\cite{Jones01}.  Current NMR experiments, which operate with
initial states that are highly mixed thermal states, use a technique
called pseudo-pure-state synthesis to process the initial thermal state
and thereby to simulate pure-state quantum information processing. This
technique suffers from an exponential loss of signal strength as the
number of qubits per molecule increases and thus is not scalable. There
is a different technique for processing the initial thermal state,
called algorithmic cooling~\cite{Schulman99}, which pumps entropy from
a subset of qubits into the remaining qubits, leaving the special
subset in a pure state and the remaining qubits maximally mixed.
Algorithmic cooling provides an in-principle method for making
liquid-state NMR---or any qubit system that begins in a thermal
state---scalable, in essence by providing an efficient algorithmic
method for cooling a subset of the initially thermal qubits to a pure
state, thereby satisfying DiVincenzo's second criterion.

Knill and Laflamme~\cite{kl98} proposed a related mixed-state
computational model, which they called DQC1, in which there is just one
initial pure qubit, along with $n$ qubits in the maximally mixed state.
Although provably less powerful than a pure-state quantum
computer~\cite{Ambainis00}, DQC1 can perform some computational tasks
efficiently for which there are no known polynomial time classical
algorithms.  In particular, a DQC1 quantum circuit can be used to
evaluate, with fixed accuracy independent of $n$, the normalized trace,
$\tr(U_n)/2^n$, of any $n$-qubit unitary operator $U_n$ that can be
implemented efficiently in terms of quantum
gates~\cite{kl98,Laflamme02}.  In Sec.~\ref{S:classical} we consider
briefly whether there might be efficient classical algorithms for
estimating the normalized trace, and we conclude that this is unlikely.
The efficient quantum algorithm for estimating the normalized trace
provides an exponential speedup over the best known classical algorithm
for simulations of some quantum processes~\cite{pklo04,elpc04}.  Knill
and Laflamme referred to the power of this mixed-state computational
model as the ``power of one qubit.''

Study of the power of one qubit is motivated partly by NMR experiments,
but our primary motivation in this paper is to investigate the role of
entanglement in quantum computation, using DQC1 as a theoretical test
bed for the investigation.  For pure-state quantum computers, Jozsa and
Linden \cite{Josza99} have shown that exponential speedup over a
classical computer requires that entanglement not be restricted to
blocks of qubits of fixed size as problem size increases. Entanglement
that increases with problem size is thus a necessary prerequisite for
the exponential speedup achieved by a pure-state quantum computer.  On
the other hand, the Gottesman-Knill theorem \cite{N&C}, demonstrates
that global entanglement is far from sufficient for exponential
speedup.  While this means that the role of entanglement is not
entirely understood for pure-state quantum computers, far less is known
about the role of entanglement in mixed-state quantum computers. When
applied to mixed-state computation, the Jozsa-Linden proof does not
show that entanglement is a requirement for exponential speedup.
Indeed, it has not previously been shown that there is any entanglement
in the DQC1 circuits that provide an exponential speedup over classical
algorithms.

The purpose of this paper is to investigate the existence of and amount
of entanglement in the DQC1 circuit that is used to estimate the
normalized trace.  The DQC1 model consists of a \emph{special qubit\/}
(qubit~0) in the initial state $|0\rangle\langle0|={1\over2}(I_1+Z)$,
where $Z$ is a Pauli operator, along with $n$ other qubits in the
completely mixed state, $I_n/2^n$, which we call the \emph{unpolarized
qubits}.  The circuit consists of a Hadamard gate on the special qubit
followed by a controlled unitary on the remaining
qubits~\cite{Laflamme02}: \vspace{-1em}
\begin{equation}
\label{E:circuit}
\Qcircuit @C=.5em @R=-.5em {
    & \lstick{\ket{0}\!\bra{0}} & \gate{H} & \ctrl{1} & \meter & \push{\rule{0em}{4em}} \\
    & & \qw & \multigate{4}{U_n} & \qw & \qw \\
    & & \qw & \ghost{U_n} & \qw & \qw \\
    \lstick{\mbox{$I_n/2^n$}} & & \qw & \ghost{U_n} & \qw & \qw \\
    & & \qw & \ghost{U_n} & \qw & \qw \\
    & & \qw & \ghost{U_n} & \qw & \qw \gategroup{2}{2}{6}{2}{.6em}{\{}
}
\end{equation}
After these operations, the state of the $n+1$ qubits becomes
\begin{equation}
\rho_{n+1}=
{1\over2N}
\Bigl(|0\rangle\langle0|\otimes I_n+|1\rangle\langle1|\otimes I_n
+|0\rangle\langle1|\otimes U_n^\dagger
+|1\rangle\langle0|\otimes U_n\Bigr)
= \frac{1}{2N}
    \begin{pmatrix}
        I_n & \, U_n^\dag \\
        \, U_n & I_n
    \end{pmatrix} \;,
\label{E:rhoout}
\end{equation}
where $N=2^n$.  The information about the normalized trace of $U_n$ is
encoded in the expectation values of the Pauli operators $X$ and $Y$ of
the special qubit, i.e., $\langle X\rangle={\rm Re}[\tr(U_n)]/2^n$ and
$\langle Y\rangle=-{\rm Im}[\tr(U_n)]/2^n$.

To read out the desired information, say, about the real part of the
normalized trace, one runs the circuit repeatedly, each time measuring
$X$ on the special qubit at the output.  The measurement results are
drawn from a distribution whose mean is the real part of the normalized
trace and whose variance is bounded above by 1.  After $L$ runs, one
can estimate the real part of the normalized trace with an accuracy
$\epsilon\sim1/\sqrt L$.  Thus, to achieve accuracy $\epsilon$ requires
that the circuit be run $L\sim1/\epsilon^2$ times. More precisely, what
we mean by estimating with fixed accuracy is the following: let $P_e$
be the probability that the estimate is farther from the true value
than $\epsilon\,$; then the required number of runs is
$L\sim\ln(1/P_e)/\epsilon^2$.  That the number of runs required to
achieve a fixed accuracy does not scale with number of qubits and
scales logarithmically with the error probability is what is meant by
saying that the DQC1 circuit provides an efficient method for
estimating the normalized trace.

Throughout much of our analysis, we use a generalization of the DQC1
circuit, in which the initial pure state of the special qubit is
replaced by the mixed state ${1\over2}(I_1+\alpha Z)$,
which has polarization $\alpha$,
\begin{equation}
\label{E:circuitalpha}
\Qcircuit @C=.5em @R=-.5em {
    & \lstick{{1\over2}(I_1+\alpha Z)}
        & \gate{H} & \ctrl{1} & \meter & \push{\rule{0em}{4em}} \\
    & & \qw & \multigate{4}{U_n} & \qw & \qw \\
    & & \qw & \ghost{U_n} & \qw & \qw \\
    \lstick{\mbox{$I_n/2^n$}} & & \qw & \ghost{U_n} & \qw & \qw \\
    & & \qw & \ghost{U_n} & \qw & \qw \\
    & & \qw & \ghost{U_n} & \qw & \qw \gategroup{2}{2}{6}{2}{.6em}{\{}
}
\end{equation}
giving an overall initial state
\begin{equation}
\rho_i={1\over2N}(I_1+\alpha Z)\otimes I_n
={1\over2N}\!\left[I_{n+1}+
\alpha
\begin{pmatrix}
I_n&0\\
0&-I_n
\end{pmatrix}
\right]\;.
\end{equation}
We generally assume that $\alpha\ge0$, except where we explicitly note
otherwise.  After the circuit is run, the system state becomes
\begin{equation}
\rho_{n+1}(\alpha)=
{1\over2N}\!\left[I_{n+1}+
\alpha
\begin{pmatrix}
0&U_n^\dag\\
U_n&0
\end{pmatrix}
\right]=
\frac{1}{2N}
    \begin{pmatrix}
        I_n &  \alpha U_n^\dag \\
        \alpha U_n & I_n
    \end{pmatrix} \;.
\label{E:rhooutalpha}
\end{equation}
The effect of subunity polarization is to reduce the expectation
values of $\langle X\rangle$ and $\langle Y\rangle$ by a factor of
$\alpha$, thereby making it more difficult to estimate the normalized
trace.  Specifically, the number of runs required to estimate the
normalized trace becomes $L\sim\ln(1/P_e)/\alpha^2\epsilon^2$.  Reduced
polarization introduces an additional overhead, but as long as the
special qubit has nonzero polarization, the model still provides an
efficient estimation of the normalized trace.  What we are dealing with
is really the ``power of even the tiniest fraction of a qubit.''

For $n+1$ qubits, \emph{all\/} states contained in a ball of radius
$r_{n+1}$ centered at the completely mixed state are
separable~\cite{b99,gb03} (distance is measured by the Hilbert-Schmidt
norm).  Unitary evolution leaves the distance from the completely mixed
state fixed, so at all times during the circuit~(\ref{E:circuitalpha}),
the system state is a fixed distance
$\sqrt{\tr(\rho_i-I_{n+1}/2N)^2}=\alpha 2^{-(n+1)/2}$ from the
completely mixed state.  This suggests that with $\alpha$ small enough,
there might be an exponential speedup with demonstrably separable
states.  This suggestion doesn't pan out, however, because the radius
of the separable ball decreases exponentially faster than
$2^{-(n+1)/2}$.  The best known lower bound on $r_n$ is $2\times
6^{-n/2}$~\cite{gb04}; for the system state to be contained in a ball
given by this lower bound, we need $\alpha\le 2\times 3^{-(n+1)/2}$.
The exponential decrease of $\alpha$ means that an exponentially
increasing number of runs is required to estimate the normalized trace
with fixed accuracy.  More to the point, the possibility that the
actual radius of the separable ball might decrease slowly enough to
avoid an exponential number of runs is ruled out by the existence of a
family of $n$-qubit entangled states found by D\"ur~\emph{et
al.}~\cite{dur99}, which establishes an upper bound on $r_n$ that goes
as $2\times 2^{-n}$ for large $n$, implying that $\alpha\le 2\times
2^{-(n+1)/2}$ if the system state is to be in the ball given by the
upper bound.  These considerations do not demonstrate the impossibility
of an exponential speedup using separable states, but they do rule out
the possibility of finding such a speedup within the maximal separable
ball about the completely mixed state.

We are thus motivated to look for entanglement in states of the
form~(\ref{E:rhooutalpha}), for at least some unitary operators $U_n$.
Initial efforts in this direction are not encouraging.  It is clear
from the start that the marginal state of the $n$ unpolarized qubits
remains completely mixed, so these qubits are not entangled among
themselves. Moreover, in the state~(\ref{E:rhooutalpha}), as was shown
in Ref.~\cite{pklo04}, the special qubit is unentangled with the $n$
unpolarized qubits, no matter what $U_n$ is used.  To see this, one
plugs the eigendecomposition of the unitary, $U_n=\sum_j
e^{i\phi_j}|e_j\rangle\langle e_j|$, into the expression for
$\rho_{n+1}(\alpha)$.  This gives a separable decomposition
\begin{equation}
\rho_{n+1}(\alpha)=
{1\over2N}\sum_j
(|a_j\rangle\langle a_j|+|b_j\rangle\langle b_j|)
\otimes|e_j\rangle\langle e_j|\;,
\end{equation}
where $|a_j\rangle=\cos\theta|0\rangle+e^{i\phi_j}\sin\theta|1\rangle$
and $|b_j\rangle=\sin\theta|0\rangle+e^{i\phi_j}\cos\theta|1\rangle$,
with $\sin2\theta=\alpha$.

No entanglement of the special qubit with the rest and no entanglement
among the rest---where then are we to find any entanglement?  We look
for entanglement relative to other divisions of the qubits into two
parts.  In such bipartite divisions the special qubit is grouped with a
subset of the unpolarized qubits.  To detect entanglement between the
two parts, we use the Peres-Horodecki partial transpose
criterion~\cite{p96,hhh96}, and we quantify whatever entanglement we
find using a closely related entanglement monotone which we call the
\emph{multiplicative negativity\/}~\cite{Vidal02}.  The Peres-Horodecki
criterion and the multiplicative negativity do not reveal all
entanglement---they can miss what is called bound entanglement---but we
are nonetheless able to demonstrate the existence of entanglement in
states of the form~(\ref{E:rhoout}) and~(\ref{E:rhooutalpha}). For
convenience, we generally refer to the multiplicative negativity simply
as the negativity.  The reader should note, as we discuss in
Sec.~\ref{S:negativity}, that the term ``negativity'' was originally
applied to an entanglement measure that is closely related to, but
different from the multiplicative negativity.

The amount of entanglement depends, of course, on the unitary
operator~$U_n$ and on the bipartite division.  We present three results
in this regard.  First, in Sec.~\ref{S:examples}, we construct a family
of unitaries $U_n$ such that for $\alpha>1/2$, $\rho_{n+1}(\alpha)$ is
entangled for all bipartite divisions that put the first and last
unpolarized qubits in different parts, and we show that for all such
divisions, the negativity is $(2\alpha+3)/4$ for $\alpha\ge1/2$ ($5/4$
for $\alpha=1$), independent of~$n$.  Second, in Sec.~\ref{S:random},
we present numerical evidence that the state $\rho_{n+1}$ of
Eq.~(\ref{E:rhoout}) is entangled for typical unitaries, i.e., those
created by random quantum circuits.  For $n+1=5,\ldots,10$, we find
average negativities between 1.155 and just above 1.16 for the
splitting that puts $\left\lfloor n/2\right\rfloor$ of the unpolarized
qubits with qubit~0.  Third, in Sec.~\ref{S:bounds}, we show that for
all unitaries and all bipartite divisions of the $n+1$ qubits, the
negativity of $\rho_{n+1}(\alpha)$ is bounded above by the constant
$\sqrt{1+\alpha^2}$ ($\sqrt2\simeq1.414$ for $\alpha=1$), independent
of $n$.  Thus, when $n$ is large, the negativity achievable by the DQC1
circuit~(\ref{E:circuit}) becomes a vanishingly small fraction of the
maximum negativity, $\sim2^{n/2}$, for roughly equal bipartite
divisions.

The layout of the paper is as follows.  In Sec.~\ref{S:classical} we
examine the classical problem of estimating the normalized trace of a
unitary.  In Sec.~\ref{S:negativity} we review pertinent properties of
the negativity before applying it to obtain our three key results in
Secs.~\ref{S:examples}-\ref{S:bounds}.  We conclude in
Sec.~\ref{S:conclusion} and prove a brief Lemma in an Appendix.
Throughout we use $\breve A$ to stand for the partial transpose of an
operator $A$ relative to a particular bipartite tensor-product
structure, and we rely on context to make clear which bipartite
division we are using at any particular point in the paper.

\section{Classical Evaluation of the Trace}\label{S:classical}

In this section we outline briefly a classical method for evaluating the
trace of a unitary operator that can be implemented efficiently in terms
of quantum gates, and we indicate why this appears to be a problem that
is exponentially hard in the number of qubits.

The trace of a unitary matrix $U_n\equiv U$ is the sum over the
diagonal matrix elements of~$U$:
\begin{equation}
\label{E:sum1}
\tr(U)=\sum_{\ba} \bra{\ba}U\ket{\ba}\;.
\end{equation}
Here $\ba$ is a bit string that specifies a computational-basis state
of the $n$ qubits.  By factoring $U$ into a product of elementary gates
from a universal set and inserting a resolution of the identity between
all the gates, we can write $\tr(U)$ as a sum over the amplitudes of
Feynman paths.  A difficulty with this approach is that the sum must be
restricted to paths that begin and end in the same state.  We can
circumvent this difficulty by preceding and succeeding $U$ with a
Hadamard gate on all the qubits.  This does not change the trace, but
does allow us to write it as
\begin{equation}
\label{E:sum2}
\tr(U)=\sum_{\ba,\bb,\bc}
\bra{\ba}H^{\otimes n}\ket{\bb}\bra{\bb}U\ket{\bc}\bra{\bc}H^{\otimes n}\ket{\ba}
={1\over2^n}\sum_{\ba,\bb,\bc}
(-1)^{\ba\cdot(\bb+\bc)}\bra{\bb}U\ket{\bc}\;.
\end{equation}
Now if we insert a resolution of the identity between the elementary
gates, we get $\tr(U)$ written as an unrestricted sum over Feynman-path
amplitudes, with an extra phase that depends on the initial and final
states.

Following Dawson~\emph{et al.}~\cite{dhhmno05}, we consider two
universal gate sets: (i)~the Hadamard gate $H$, the $\pi/4$ gate $T$,
and the controlled-NOT gate and (ii)~$H$ and the Toffoli gate.  With
either of these gate sets, most of the Feynman paths have zero
amplitude.  Dawson~\emph{et al.}~\cite{dhhmno05} introduced a
convenient method, which we describe briefly now, for including only
those paths with nonzero amplitude.  One associates with each wire in
the quantum circuit a classical bit value corresponding to a
computational basis state.  The effect of an elementary gate is to
change, deterministically or stochastically, the bit values at its
input and to introduce a multiplicative amplitude.  The two-qubit
controlled-NOT gate changes the input control bit $x$ and target bit
$y$ deterministically to output values $x$ and $y\oplus x$, while
introducing only unit amplitudes.  Similarly, the three-qubit Toffoli
gates changes the input control bits $x$ and $y$ and target bit $z$
deterministically to $x$, $y$, and $z\oplus xy$, while introducing only
unit amplitudes.  The $T$ gate leaves the input bit value $x$ unchanged
and introduces a phase $e^{ix\pi/4}$.  The Hadamard gate changes the
input bit value $x$ stochastically to an output value $y$ and
introduces an amplitude $(-1)^{xy}/\sqrt2$.

The classical bit values trace out the allowed Feynman paths, and the
product of the amplitudes introduced at the gates gives the overall
amplitude of the path.  In our application of evaluating the
trace~(\ref{E:sum2}), a path is specified by $n$ input bit values
(which are identical to the output bit values), $n$ random bit values
introduced by the initial Hadamard gates, and $h$ random bit values
introduced at the $h$ Hadamard gates required for the implementation of
$U$.  This gives a total of $2n+h$ bits to specify a path and thus
$2^{2n+h}$ allowed paths.  We let $\bx$ denote collectively the $2n+h$
path bits.

If we apply the gate rules to a Hadamard-Toffoli circuit, the only gate
amplitudes we have to worry about are the $\pm1/\sqrt2$ amplitudes
introduced at the Hadamard gates.  There being no complex amplitudes,
the trace cannot be complex.  Indeed, for this reason, achieving
universality with the $H$-Toffoli gate set requires the use of a simple
encoding, and we assume for the purposes of our discussion that this
encoding has already been taken into account.  With all this in mind,
we can write the trace~(\ref{E:sum2}) as a sum over the allowed paths,
\begin{equation}
\tr(U)=\frac{1}{2^{n+h/2}}\sum_{\bx}(-1)^{\psi(\bx)}\;.
\end{equation}
Here $\psi(\bx)$ is a polynomial over $\mz_2$, specifically, the mod-2
sum of the products of input and output bit values at each of the
Hadamard gates.  The downside is that a string of Toffoli gates
followed by a Hadamard can lead to a polynomial that is high order in
the bit values.  As pointed out by Dawson~\emph{et
al.}~\cite{dhhmno05}, we can deal with this problem partially by
putting a pair of Hadamards on the target qubit after each Toffoli
gate, thus replacing the quadratic term in the output target bit with
two new random variables and preventing the quadratic term from
iterating to higher order terms in subsequent Toffoli gates.  In doing
so, we are left with a cubic term in $\psi(\bx)$ from the amplitude of
the first Hadamard.  The upshot is that we can always make $\psi(\bx)$
a cubic polynomial.

Notice now that we can rewrite the trace as
\begin{equation}
\tr(U)
=\frac{1}{2^{n+h/2}}\left[
\begin{pmatrix}
\mbox{number of $\bx$ such}\\ \mbox{that $\psi(\bx)=0$}
\end{pmatrix}-
\begin{pmatrix}
\mbox{number of $\bx$ such}\\ \mbox{that $\psi(\bx)=1$}
\end{pmatrix}\right]\;,
\end{equation}
thus reducing the problem of evaluating the trace exactly to counting
the number of zeroes of the cubic polynomial $\psi(\bx)$.  This is a
standard problem from computational algebraic geometry, and it is known
that counting the number of zeroes of a general cubic polynomial over
any finite field is \#{\bf P} complete \cite{ek90}.  It is possible
that the polynomials that arise from quantum circuits have some special
structure that can be exploited to give an efficient algorithm for
counting the number of zeroes, but in the absence of such structure,
there is no efficient classical algorithm for computing the trace
exactly unless the classical complexity hierarchy collapses and
\emph{all\/} problems in \#{\bf P} are efficiently solvable on a
classical computer.

Of course, it is not our goal to compute the trace exactly, since the
quantum circuit only provides an efficient method for estimating the
normalized trace to fixed accuracy.  This suggests that we should
estimate the normalized trace by sampling the amplitudes of the allowed
Feynman paths.  The normalized trace,
\begin{equation}
{\tr(U)\over2^n}=\frac{1}{2^{2n+h}}\sum_{\bx}2^{h/2}(-1)^{\psi(\bx)}\;,
\end{equation}
which lies between $-1$ and $+1$, can be regarded as the average of
$2^{2n+h}$ quantities whose magnitude, $2^{h/2}$, is exponentially
large in the number of Hadamard gates.  To estimate the average with
fixed accuracy requires a number of samples that goes as $2^h$,
implying that this is not an efficient method for estimating the
normalized trace.  The reason the method is not efficient is pure
quantum mechanics, i.e., that the trace is a sum of amplitudes, not
probabilities.

If we apply the gate rules to a Hadamard-$T$-controlled-NOT circuit,
the bit value on each wire in the circuit is a mod-2 sum of appropriate
bit values in $\bx$, but now we have to worry about the amplitudes
introduced by the Hadamard and $T$ gates.  The trace~(\ref{E:sum2})
can be written as
 \begin{equation}
  \tr(U)={1\over2^{n+h/2}}
  \sum_{\bx}e^{i(\pi/4)\chi(\bx)}(-1)^{\phi(\bx)}\;.
  \label{E:sum3}
 \end{equation}
Here $\phi(\bx)$ is a polynomial over $\mz_2$, obtained as the mod-2
sum of the products of input and output bit values at each of the
Hadamard gates.  Since the output value is a fresh binary variable and
the input value is a mod-2 sum of bit values in $\bx$, $\phi(\bx)$ is a
purely quadratic polynomial over $\mz_2$.  The function $\chi(\bx)$ is
a mod-8 sum of the input bit values to all of the $T$ gates.  Since
these input bit values are mod-2 sums of bit values in $\bx$,
$\chi(\bx)$ is linear in bit values, but with an unfortunate mixture of
mod-2 and mod-8 addition.  We can get rid of this mixture by preceding
each $T$ gate with a pair of Hadamards, thus making the input to the
every $T$ gate a fresh binary variable.  With this choice, $\chi(\bx)$
becomes a mod-8 sum of appropriate bit values from $\bx$.

We can rewrite the sum~(\ref{E:sum3}) in the following way:
\begin{equation}
 \tr(U)={1\over2^{n+h/2}}
 \sum_{j=0}^7 e^{i(\pi/4)j} \left[
\begin{pmatrix}
\mbox{number of $\bx$ such that}\\
\mbox{$\chi(\bx)=j$ and $\phi(\bx)=0$}
\end{pmatrix}
-
\begin{pmatrix}
\mbox{number of $\bx$ such that}\\
\mbox{$\chi(\bx)=j$ and $\phi(\bx)=1$}
\end{pmatrix}
\right].
\end{equation}
Thus the problem now reduces to finding simultaneous (binary) solutions
to the purely quadratic $\mz_2$ polynomial $\phi(\bx)$ and the purely
linear $\mz_8$ polynomial $\chi(\bx)$.  One has to be careful here to
note that we are only interested in binary solutions, so we are not
solving $\chi(\bx)=j$ over all values in $\mz_8$.  The number of
solutions of a purely quadratic polynomial over $\mz_2$ can be obtained
trivially~\cite{ek90}, but the constraint over $\mz_8$ means that one
must count the number of solutions over a mixture of a field and a
ring.  The complexity class for this problem is not known, but given
the equivalence to counting the number of solutions of a cubic
polynomial over $\mz_2$, it seems unlikely that there is an efficient
classical algorithm.  Moreover, an attempt to estimate the normalized
trace by sampling allowed paths obviously suffers from the problem
already identified above.

\section{Properties of negativity}
\label{S:negativity}

In this section we briefly review properties of negativity as an
entanglement measure, focusing on those properties that we need in the
subsequent analysis (for a thorough discussion of negativity, see
Ref.~\cite{Vidal02}).

Let $A$ be an operator in the joint Hilbert space of two systems,
system~1 of dimension $d_1$ and system~2 of dimension $d_2$.  The
partial transpose of $A$ with respect to an orthonormal basis of
system~2 is defined by taking the transpose of the matrix elements of
$A$ with respect to the system-2 indices.  A partial transpose can also
be defined with respect to any basis of system~1.  Partial
transposition preserves the trace, and it commutes with taking the
adjoint.

The operator that results from partial transposition depends on which
basis is used to define the transpose, but these different partial
transposes are related by unitary transformations on the transposed
system and thus have the same eigenvalues and singular values.
Moreover, partial transposition on one of the systems is related to
partial transposition on the other by an overall transposition, which
also preserves eigenvalues and singular values.  Despite the
nonuniqueness of the partial transpose, we can talk meaningfully about
its invariant properties, such as its eigenvalues and singular values.
Similar considerations show that the eigenvalues and singular values
are invariant under local unitary transformations.

The singular values of an operator $O$ are the eigenvalues of
$\sqrt{O^\dag O}\equiv|O|$ (or, equivalently, of $\sqrt{O
O^\dag}$).  Any operator has a polar decomposition $O=T|O|$, where
$T$ is a unitary operator.  Writing $|O|=W^\dag SW$, where $W$ is
the unitary that diagonalizes $|O|$ and $S$ is the diagonal matrix of
singular values, we see that any operator can be written as $O=VSW$,
where $V=TW^\dag$ and $W$ are unitary operators.

We denote a partial transpose of $A$ generically by $\breve A$.  We
write the eigenvalues of $\breve A$ as $\lambda_j(\breve A)$ and denote
the singular values by $s_j(\breve A)$.  If $A$ is Hermitian, so is
$\breve A$, and the singular values of $\breve A$, i.e., the
eigenvalues of $|\breve A|$, are the magnitudes of the eigenvalues,
i.e, $s_j(\breve A)=|\lambda_j(\breve A)|$.

If a joint density operator~$\rho$ of systems 1 and 2 is separable, its
partial transpose $\breve\rho$ is a positive operator.  This gives the
Peres-Horodecki entanglement criterion~\cite{p96,hhh96}: if
$\breve\rho$ has a negative eigenvalue, then $\rho$ is entangled (the
converse is not generally true).  The magnitude of the sum of the
negative eigenvalues of the partial transpose, denoted by
\begin{equation}
\sN(\rho)\equiv-\sum_{\lambda_j(\breve\rho)<0}\lambda_j(\breve\rho)\;,
\end{equation}
is a measure of the amount of entanglement.  Partial transposition
preserves the trace, so $\tr(\breve\rho)=1$, from which we get
\begin{equation}
1+2\sN(\rho)=\sum_j|\lambda_j(\breve\rho)|=
\tr|\breve\rho|\equiv\sM(\rho)\;,
\end{equation}
where $\sM(\rho)$ is a closely related entanglement measure.  The
quantity $\sN(\rho)$ was originally called the
\emph{negativity\/}~\cite{Vidal02}; we can distinguish the two measures
by referring to $\sM(\rho)$ as the \emph{multiplicative negativity}, a
name that emphasizes one of its key properties and advantages over
$\sN(\rho)$.  In this paper, however, we use the multiplicative
negativity exclusively and so refer to it simply as ``the negativity''.

The negativity $\sM(\rho)$ equals one for separable states, and it is
an entanglement monotone~\cite{Vidal02}, meaning that (i)~it is a
convex function of density operators and (ii)~it does not increase
under local operations and classical communication.  The negativity has
the property of being multiplicative in the sense that the $\sM$ value
for a state that is a product of states for many pairs of systems is
the product of the $\sM$ values for each of the pairs.  By the same
token, $\log\sM(\rho)$, called the \emph{log-negativity\/}, is
additive, but the logarithm destroys convexity so the log-negativity is
not an entanglement monotone~\cite{Vidal02}.  For another point of view
on the monotonicity of the log-negativity, see Ref.~\cite{Plenio05}.

The minimum value of the negativity is one, but we need to know the
maximum value to calibrate our results.  Convexity guarantees that the
maximum value is attained on pure states.  We can find the
maximum~\cite{SLee03} by considering the Schmidt decomposition of a
joint pure state of systems~1 and 2,
\begin{equation}
|\psi\rangle=\sum_{j=1}^d\sqrt{\mu_j}|j,j\rangle\;,
\end{equation}
where $d=\min(d_1,d_2)$.  Taking the partial transpose of $\rho$
relative to the Schmidt basis of system~2 gives
\begin{equation}
\breve\rho
=\sum_{j,k=1}^d\sqrt{\mu_j\mu_k}|j,k\rangle\langle k,j|\;,
\end{equation}
with eigenvectors and eigenvalues
\begin{eqnarray}
|j,j\rangle\;,&&\mbox{eigenvalue $\mu_j$,}\nonumber\\
{1\over\sqrt2}(|j,k\rangle\pm|k,j\rangle)\;,
&&\mbox{eigenvalue $\pm\sqrt{\mu_j\mu_k}$,}\quad j<k.
\end{eqnarray}
This gives a negativity
\begin{equation}
\sM(\psi)=
1+2\sum_{j<k}\sqrt{\mu_j\mu_k}
=\sum_{j,k=1}^d\sqrt{\mu_j\mu_k}
=\biggl(\sum_{j=1}^d\sqrt{\mu_j}\biggr)^2\;.
\end{equation}
The concavity of the square root implies $\sum_j\sqrt{\mu_j}\le\sqrt
d$, with equality if and only if $\mu_j=1/d$ for all $j$, i.e.,
$|\psi\rangle$ is maximally entangled. We end up with
\begin{equation}
1\le\sM(\rho)\le d\;.
\end{equation}

The negativity is the sum of the singular values of
$\breve\rho$.  For states of the form we are interested in, given by
Eq.~(\ref{E:rhooutalpha}), the negativity is determined by the
singular values of the partial transpose of the unitary operator~$U_n$.
To see this, consider any bipartite division of the qubits. Performing
the partial transpose on the part that does not include the special
qubit, we have
\begin{equation}
\breve\rho_{n+1}(\alpha)=
\frac{1}{2N}
    \begin{pmatrix}
        I_n &  \alpha\breve U_n^\dag \\
        \alpha\breve U_n & I_n
    \end{pmatrix} \;,
\label{E:breverhooutalpha}
\end{equation}
where $\breve U_n$ is the partial transpose of $U_n$ relative to the
chosen bipartite division.  Notice that if we make our division between
the special qubit and all the rest, then $\breve U_n=U_n^T$ is a
unitary operator, and $\breve\rho_{n+1}(\alpha)$ is the quantum state
corresponding to using $U_n^T$ in the circuit~(\ref{E:circuitalpha});
this shows that for this division, the negativity is 1, consistent with
our earlier conclusion that the special qubit is not entangled with the
other qubits.  For a general division, we know there are unitaries $V$
and $W$ such that $\breve U_n=VSW$, where $S$ is the diagonal matrix of
singular values $s_j(\breve U_n)$.  This allows us to write
\begin{equation}
\breve\rho_{n+1}(\alpha)=
    \begin{pmatrix}
        W^\dagger & 0 \\
        0 & V
    \end{pmatrix}
    \frac{1}{2N}
    \begin{pmatrix}
        I_n &  \alpha S \\
        \alpha S & I_n
    \end{pmatrix}
    \begin{pmatrix}
        W & 0 \\
        0 & V^\dagger
    \end{pmatrix}\;,
\end{equation}
showing that $\breve\rho_{n+1}(\alpha)$ is a unitary transformation
away from the matrix in the middle and thus has the same eigenvalues.
The block structure of the middle matrix makes it easy to find these
eigenvalues, which are given by $[1\pm\alpha s_j(\breve U_n)]/2N$.
This allows us to put the negativity in the form
\begin{equation}
\sM\bigl(\rho_{n+1}(\alpha)\bigr)={1\over2N}
\sum_{j=1}^N|1+\alpha s_j(\breve U_n)|+|1-\alpha s_j(\breve U_n)| =
{1 \over N} \sum_{j=1}^N \max\bigl(\abs{\alpha} s_j(\breve U_n), 1\bigr)\;,
\label{E:Msingular}
\end{equation}
which is valid for both positive and negative values of $\alpha$.  An
immediate consequence of Eq.~(\ref{E:Msingular}) is that
$\sM\bigl(\rho_{n+1}(\alpha)\bigr)=\sM\bigl(\rho_{n+1}(-\alpha)\bigr)$,
as one would expect.  Since $\rho_{n+1}(\alpha)$ is a mixture of
$\rho_{n+1}(+1)=\rho_{n+1}$ and $\rho_{n+1}(-1)$, convexity tells us
immediately that
$\sM\bigl(\rho_{n+1}(\alpha)\bigr)\le\sM\bigl(\rho_{n+1}\bigr)$, i.e.,
that a mixed input for the special qubit cannot increase the negativity
over that for a pure input.  More generally, we have that the
negativity cannot decrease at any point as $\alpha$ increases from 0
to~1.

\section{Entanglement in the DQC1 Circuit} \label{S:examples}

In this section, we construct a family of unitaries $U_n$ that produce
global entanglement in the DQC1 circuit~(\ref{E:circuit}).  For
$\alpha=1$, the negativity produced by this family is equal to $5/4$,
independent of $n$, for all bipartite divisions that put the first and
last unpolarized qubits in different parts.  We conjecture that this is
the maximum negativity that can be achieved in a circuit of the
form~(\ref{E:circuit}).

Before the measurement, the output state of the
circuit~(\ref{E:circuitalpha}) is given by Eq.~(\ref{E:rhooutalpha}).
To construct the unitaries $U_n$, we first introduce a two-qubit
unitary matrix
\begin{equation}
\label{E:U2}
    U_{2} \equiv \begin{pmatrix}
        A_1 & C_1 \\
        D_1 & B_1 \\
    \end{pmatrix} \;,
\end{equation}
where $A_1$, $B_1$, $C_1$, and $D_1$ are single-qubit ($2\times2$)
matrices that must satisfy $A_1^\dagger A_1+D_1^\dagger D_1=
B_1^\dagger B_1+C_1^\dagger C_1=I_1$ and $A_1^\dagger C_1+D_1^\dagger
B_1=0$ to ensure that $U_2$ is unitary.  The $n$-qubit unitary $U_n$
is then defined by
\begin{eqnarray}
    U_n &\equiv&
    \begin{pmatrix}
        I_{n-2} \otimes A_1 & X_{n-2} \otimes C_1 \\
        X_{n-2} \otimes D_1 & I_{n-2} \otimes B_1
    \end{pmatrix} \nonumber \\
    &=&|0\rangle\langle0|\otimes I_{n-2}\otimes A_1
    +|1\rangle\langle1|\otimes I_{n-2}\otimes B_1 \nonumber \\
    &&\phantom{|}
    +|0\rangle\langle1|\otimes X_{n-2}\otimes C_1
    +|1\rangle\langle0|\otimes X_{n-2}\otimes D_1
    \;.
    \label{E:Un}
\end{eqnarray}
Here we use $X_1$, $Y_1$, and $Z_1$ to denote single-qubit Pauli
operators.  A subscript $k$ on the identity operator or a Pauli operator
denotes a tensor product in which that operator acts on each of $k$
qubits.  If we adopt the convention that $X_0 = I_0 = 1$, then $U_n$
reduces to $U_2$ when $n=2$.  It is easy to design a quantum circuit
that realizes $U_n$.  The structure of the circuit is illustrated by
the case of $U_4$:
\begin{equation}
\label{E:Ucircuit}
    \Qcircuit @C=.5em @R=0.5em {
    & \lstick{\mbox{1st qubit}} & \ctrl{2} & \ctrl{3} & \multigate{1}{U_2} & \ctrl{3} & \ctrl{2} & \qw \\
    & \lstick{\mbox{4th qubit}} & \qw      & \qw      & \ghost{U_n}        & \qw    & \qw & \qw \\
    & \lstick{\mbox{2nd qubit}} & \targ    & \qw      & \qw                & \qw      & \targ & \qw \\
    & \lstick{\mbox{3rd qubit}} & \qw      & \targ    & \qw                & \targ    & \qw & \qw                    }
\end{equation}
In general, the two-qubit unitary $U_2$, acting on the first and last
qubits, is bracketed by controlled-NOT gates from the first qubit,
acting as control, to each of the other qubits, except the last, as
targets.

Because $I_1$ and $X_1$ are invariant under transposition, it is clear
from the form of $U_n$ that in the state~(\ref{E:rhooutalpha}), all
qubits, except~0, 1, and~$n$, are invariant under transposition.  We
can use this fact to find the negativity for all bipartite divisions.
First consider any bipartite division that puts qubits 1 and $n$ in the
same part.  There are two possibilities.  If the special qubit is in
the same part as qubits~1 and $n$, then partial transposition on the
other part leaves $\rho_{n+1}(\alpha)$ unchanged, so the negativity
is~1.  If the special qubit is not in the same part as~1 and $n$, then
partial transposition on the part that includes~1 and $n$ is the same
as partial transposition of all the unpolarized qubits, a case we
already know to have negativity equal to 1.  We conclude that any
bipartite division that puts~1 and $n$ in the same part has negativity
equal to 1.

Turn now to bipartite divisions that put qubits~1 and $n$ in different
parts.  There are two cases to consider: (i)~the special qubit is in
the same part as qubit~1, and (ii)~the special qubit is in the same part
as qubit~$n$.  In case~(i), partial transposition of the part that
contains qubit~$n$ gives
\begin{equation}
\breve\rho_{n+1}(\alpha)={1\over2N}
    \begin{pmatrix}
        I_n                 & \alpha \breve U_n^\dagger \\
        \alpha \breve U_n   & I_n
    \end{pmatrix}
\qquad\mbox{with}\qquad
\breve U_n=
\begin{pmatrix}
        I_{n-2} \otimes A_1^T & X_{n-2} \otimes C_1^T \\
        X_{n-2} \otimes D_1^T & I_{n-2} \otimes B_1^T
    \end{pmatrix}\;.
\end{equation}
In case~(ii), partial transposition of the part that contains qubit~1
gives
\begin{equation}
\breve\rho_{n+1}(\alpha)={1\over2N}
    \begin{pmatrix}
        I_n                 & \alpha \breve U_n^\dagger \\
        \alpha \breve U_n   & I_n
    \end{pmatrix}
\qquad\mbox{with}\qquad
\breve U_n=
\begin{pmatrix}
        I_{n-2} \otimes A_1 & X_{n-2} \otimes D_1 \\
        X_{n-2} \otimes C_1 & I_{n-2} \otimes B_1
    \end{pmatrix}\;.
\end{equation}

The basic structure of $\breve\rho_{n+1}(\alpha)$ is the same in
both cases.  Without changing the spectrum, we can reorder the rows and
columns to block diagonalize $\breve\rho_{n+1}(\alpha)$ so that
there are $N/4$ blocks, each of which has the form
\begin{equation}
{1\over2N}
\begin{pmatrix}
    I_2                 & \alpha \breve U_2^\dagger \\
    \alpha \breve U_2   & I_2
\end{pmatrix}
={4\over N}\breve\rho_3(\alpha)\;,
\label{E:breverho2}
\end{equation}
where $\breve\rho_3(\alpha)$ is the appropriate partial transpose of
the three-qubit output state.  Thus the spectrum of
$\breve\rho_{n+1}(\alpha)$ is the same as the spectrum of
$\breve\rho_3(\alpha)$, except that each eigenvalue is reduced by a
factor of $4/N$.  In calculating the negativity, since each eigenvalue
is ($N/4$)-fold degenerate, the reduction factor of $4/N$ is cancelled
by a degeneracy factor of $N/4$, leaving us with the fundamental result
of our construction,
\begin{equation}
\sM\bigl(\rho_{n+1}(\alpha)\bigr)=\sM\bigl(\rho_3(\alpha)\bigr)\;.
\end{equation}
This applies to both cases of bipartite splittings that we are
considering, showing that all divisions have the same negativity as the
corresponding $n=2$ construction.

We now specialize to a particular choice of $U_2$ given by
\begin{equation}
A_1=
    \begin{pmatrix}
        0 & 0 \\ 0 & 1
    \end{pmatrix}\;,
\quad
B_1=
    \begin{pmatrix}
        1 & 0 \\ 0 & 0
    \end{pmatrix}\;,
\quad
C_1=
    \begin{pmatrix}
        0 & 1 \\ 0 & 0
    \end{pmatrix}\;,
\quad\mbox{and}\quad
D_1=
    \begin{pmatrix}
        0 & 0 \\ 1 & 0
    \end{pmatrix}\;.
\label{E:ABCD}
\end{equation}
For this choice, the two cases of bipartite division lead to the same
partial transpose.  The spectrum of
\begin{equation}
\breve\rho_3(\alpha)={1\over8}
\begin{pmatrix}
I_1 & 0 & \alpha A_1 & \alpha D_1 \\
0 & I_1 & \alpha C_1 & \alpha B_1  \\
\alpha A_1 & \alpha D_1 & I_1 & 0\\
\alpha C_1 & \alpha B_1 & 0 & I_1
\end{pmatrix}
\end{equation}
is
\begin{equation}
\spec(\breve\rho_3\bigl(\alpha)\bigr)=
{1\over 8}(1+2\alpha,1,1,1,1,1,1-2\alpha)\;,
\end{equation}
giving a negativity equal to 1 for $\alpha\le1/2$ and a
negativity
\begin{equation}
\sM\bigl(\rho_{n+1}(\alpha)\bigr)=\sM\bigl(\rho_3(\alpha)\bigr)=
{1\over4}(2\alpha+3)
\quad\mbox{for}\quad
\alpha\ge1/2\;.
\end{equation}
This result shows definitively that the circuit~ can produce
entanglement, at least for $\alpha>1/2$.  We stress that the
negativity achieved by this family of unitaries is independent of
$n\ge2$.

For $\alpha=1$, the negativity achieved by this family reduces to
$5/4$.  For large $n$, this amount of negativity is a vanishingly small
fraction of the maximum possible negativity, $\sim2^{n/2}$, for roughly
equal divisions of the qubits.  This raises the question whether it is
possible for other unitaries to achieve larger negativities.  A first
idea might be to find two-qubit unitaries $U_2$ that yield a higher
negativity $\sM(\rho_3)=\sM(\rho_{n+1})$ when plugged into the
construction of this section, but the bounds we find in
Sec.~\ref{S:bounds} dispose of this notion, since they show that $5/4$
is the maximum negativity that can be achieved for $n=2$.  Another
approach would be to generalize the construction of this section in a
way that is obvious from the circuit~(\ref{E:Ucircuit}), i.e., by
starting with a $k$-qubit unitary in place of the two-qubit unitary of
Eq.~(\ref{E:Ucircuit}).  Numerical investigation of the case $k=3$ has
not turned up negativities larger than $5/4$.  We conjecture that $5/4$
is the maximum negativity that can be achieved by states of the
form~(\ref{E:rhoout}).  Though we have not been able to prove this
conjecture, we show in the next section that typical unitaries for
$n+1\le10$ achieve negativities less than $5/4$ and in the following
section that the negativity is rigorously bounded by $\sqrt2$.

We stress that we are not suggesting that the construction of this
section, with $U_2$ given by Eq.~(\ref{E:ABCD}), achieves the maximum
negativity for all values of $\alpha$, for that would mean that we
believed that the negativity cannot exceed 1 for $\alpha\le1/2$, which
we do not.  Although we have not found entanglement for $\alpha\le
1/2$, we suspect there are states with negativity greater than 1 as
long as $\alpha$ is large enough that $\rho_{n+1}(\alpha)$ lies outside
the separable ball around the maximally mixed state, i.e.,
$\alpha\ge2^{(n+1)/2}r_{n+1}$.  The bound of Sec.~\ref{S:bounds} only
says that $\sM\bigl(\rho_{n+1}(\alpha)\bigr)\le\sqrt{1+\alpha^2}$, thus
allowing negativities greater than 1 for all values of $\alpha$ except
$\alpha=0$.  Moreover, since the negativity does not detect bound
entanglement, there could be entangled states that have a negativity
equal to~1.

\section{The Average Negativity of a Random Unitary} \label{S:random}

Having constructed a family of unitaries that yields a DQC1 state with
negativity $5/4$, a natural question to ask is, ``What is the
negativity of a typical state produced by the
circuit~(\ref{E:circuit})?''  To address this question, we choose the
unitary operator in the circuit~(\ref{E:circuit}) at random and
calculate the negativity.  Of course, one must first define what it
means for a unitary to be ``typical'' or ``chosen at random''.  The
natural measure for defining this is the Haar measure, which is the
unique left-invariant measure for the group ${\sf U}(N)$
\cite{Conway90}.  The resulting ensemble of unitaries is known as the
Circular Unitary Ensemble, or CUE, and it is parameterized by the
Hurwitz decomposition \cite{Hurwitz1897}.  Although this is an exact
parameterization, implementing it requires computational resources that
grow exponentially in the size of the unitary~\cite{Emerson03}.  To
circumvent this, a pseudo-random distribution that requires resources
growing polynomially in the size of the unitary was formulated and
investigated in Ref.~\cite{Emerson03}. This is the distribution from
which we draw our random unitaries, and we summarize the procedure for
completeness.

We first define a random ${\sf SU}(2)$ unitary as
\begin{equation}
    R(\theta,\phi,\chi) =
    \begin{pmatrix}
    e^{i \phi} \cos\theta & e^{i \chi} \sin\theta \\
    -e^{-i \chi} \sin\theta & e^{-i \phi} \cos\theta \\
    \end{pmatrix} \;,
\end{equation}
where $\theta$ is chosen uniformly between $0$ and $\pi/2$, and $\phi$
and $\chi$ are chosen uniformly between $0$ and $2\pi$.  A random
unitary applied to each of the $n$ qubits is then
\begin{equation}
    {\sf R} = \bigotimes_{i=1}^{n} R(\theta_i, \phi_i, \chi_i) \;,
\end{equation}
where a separate random number is generated for each variable at each
value of $i$.  Now define a mixing operator ${\sf M}$ in terms of
nearest-neighbor $Z\otimes Z$ couplings as
\begin{equation}
    {\sf M} =
    \exp\left(i \frac{\pi}{4} \sum_{j=1}^{n-1} Z^{(j)} \otimes Z^{(j+1)} \right) \;.
\end{equation}
The pseudo-random unitary is then given by
\begin{equation}
    {\sf R}_j{\sf M}{\sf R}_{j-1}\cdots{\sf M}{\sf R}_2{\sf M}{\sf R}_1 \;,
\end{equation}
where $j$ is a positive integer that depends on $n$, and each ${\sf
R}_k$ is chosen randomly as described above.  For a given $n$, the
larger $j$ is, the more accurately the pseudo-random unitary
distribution resembles the actual CUE.  From the results in
Ref.~\cite{Emerson03}, $j=40$ gives excellent agreement with the CUE
for unitary operators on at least up to 10 qubits, so this is what we
use in our calculations.

\begin{figure}[t]
\begin{center}
\includegraphics[scale=.458]{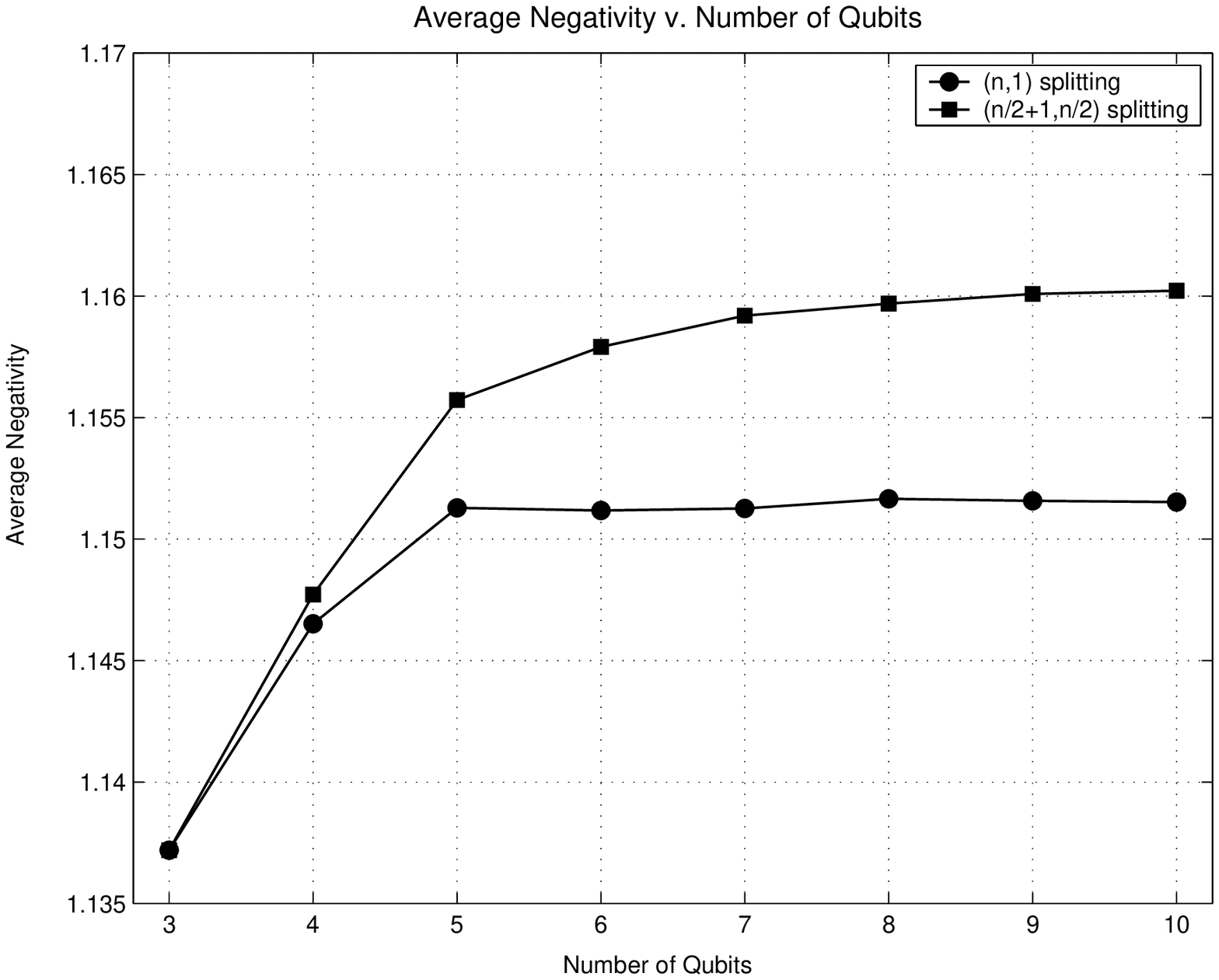}
\includegraphics[scale=.458]{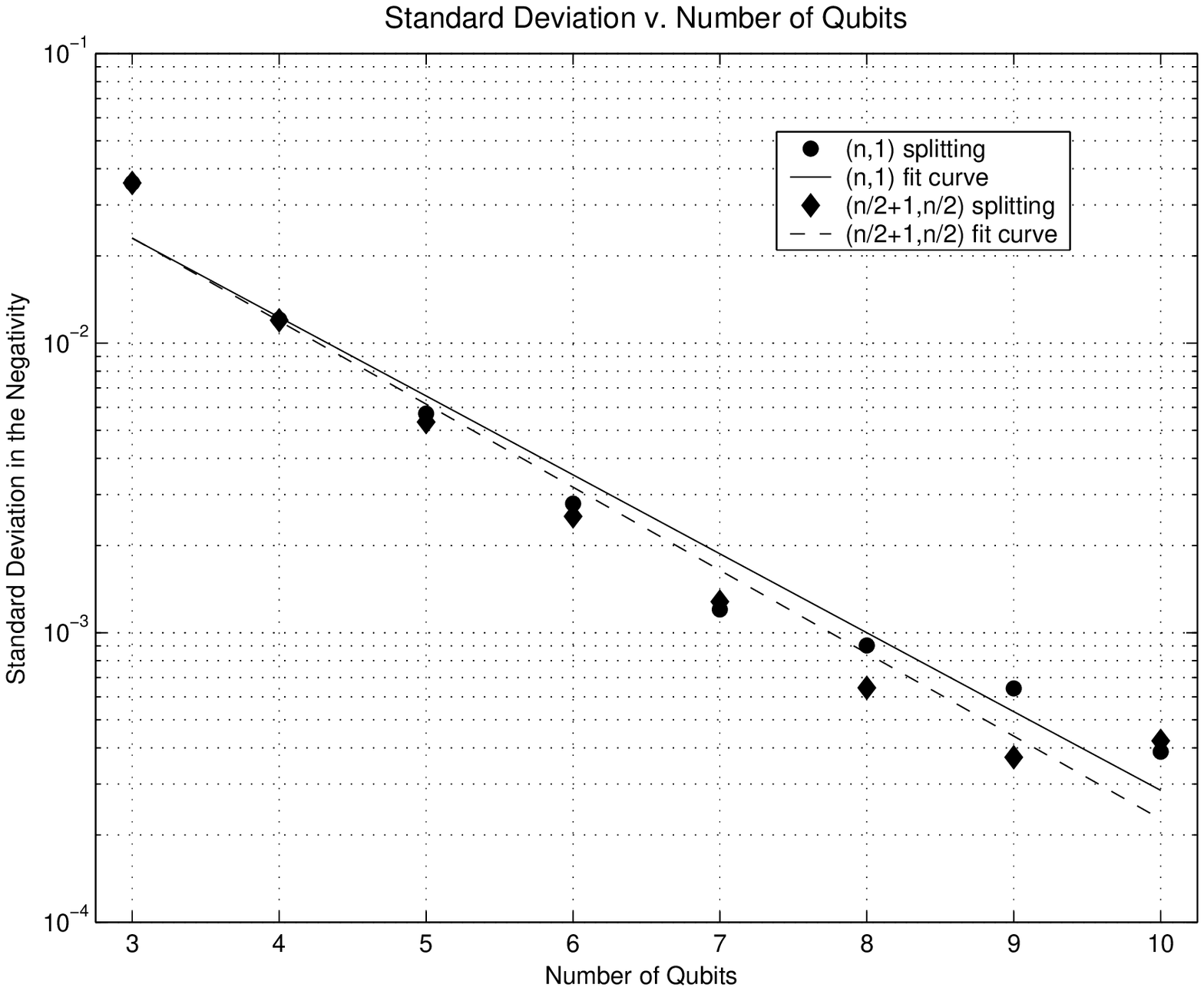}
\caption{{\bf Left}: Average negativity of the state $\rho_{n+1}$ of
Eq.~(\protect\ref{E:rhoout}) ($\alpha=1$) for a randomly chosen unitary
$U_n$ for two different bipartite splittings, $(n,1)$ and
$\left(\left\lfloor n/2 \right\rfloor +1, \left\lceil
n/2 \right\rceil\right)$.  The $(n,1)$ splitting appears to
reach an upper bound quickly, whereas the other splitting is still
rising slowly at 10 qubits.  {\bf Right}: Semi-log plot of the standard
deviation in the negativity of the randomly chosen state $\rho_{n+1}$.
The fit curves show that the standard deviation is decaying
exponentially, so that for large numbers of qubits, almost all
unitaries give the same negativity.} \label{F:random}
\end{center}
\end{figure}

\begin{figure}[t]
\begin{center}
\includegraphics[scale=.6]{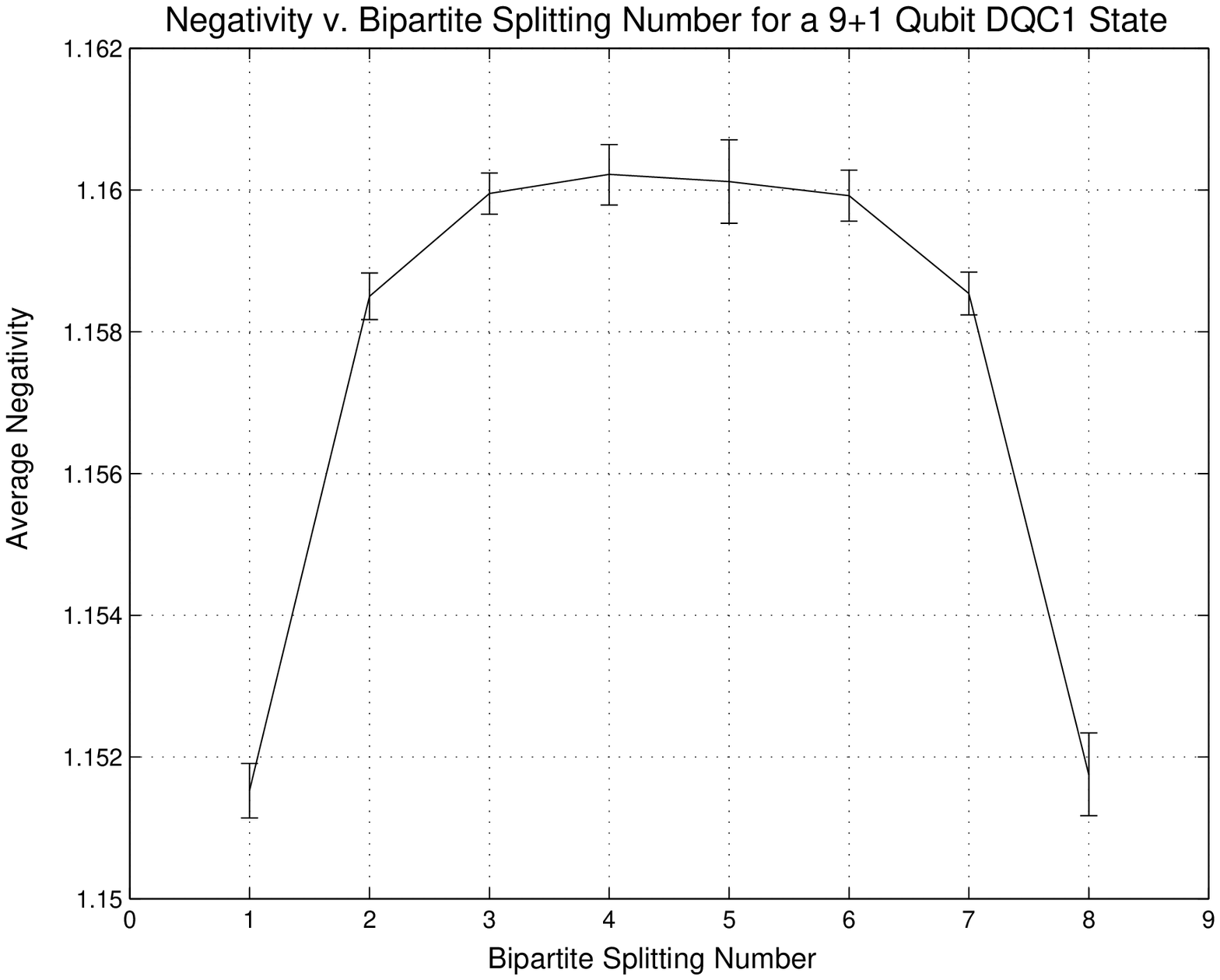}
\caption{Average negativity of the state $\rho_{10}$ of
Eq.~(\protect\ref{E:rhoout}) ($\alpha=1$) for a randomly chosen unitary
$U_9$ as a function of bipartite splitting number $k$ for bipartite
splittings $(10-k,k)$.  The error bars give the standard deviations.
The function attains a maximum when the bipartite split is made between
half of the qubits on which the unitary acts.}
\label{F:allsplits}
\end{center}
\end{figure}

Due to boundary effects, not all bipartite splittings that put $k$
unpolarized qubits in one part are equivalent. Nevertheless, we
consider only bipartite divisions that split the qubits along
horizontal lines placed at various points in the circuit of
Eq.~(\ref{E:circuit}).  We refer to the division that groups the last
$k$ qubits together as the $(n+1-k,k)$ splitting.  For $\alpha=1$, we
calculate the average negativity and standard deviation of a
pseudo-random state $\rho_{n+1}$ for two different bipartite
splittings, $(n,1)$ and $(\left\lfloor n/2 \right\rfloor +1,
\left\lceil n/2 \right\rceil )$. These results are plotted in
Fig.~\ref{F:random}.  For $n+1=5,\ldots,10$, the average negativity for
the roughly equal splitting lies between 1.135 and just above 1.16. The
standard deviation appears to converge exponentially to zero, as in
Ref.~\cite{Scott03}, a behavior that is typical of asymptotically
equivalent matrices.  In addition, for $9+1$ qubits, we calculate the
average negativity and standard deviation for all nontrivial ($k\ne n$)
bipartite splittings $(n+1-k,k)$, and the results are shown in
Fig.~\ref{F:allsplits}.

\section{Bounds on the Negativity} \label{S:bounds}

In this section, we return to allowing the special qubit in the circuit
(\ref{E:circuitalpha}) to have initial polarization~$\alpha$.  Since the
value of $n$ is either clear from context or fixed, we reduce the
notational clutter by denoting the state $\rho_{n+1}(\alpha)$ of
Eq.~(\ref{E:rhooutalpha}) as~$\rho_\alpha$.

Given a particular bipartite division, the partial transpose of
$\rho_\alpha$ with respect to the part that does not include the
special qubit is
\begin{equation}
    \breve \rho_\alpha = \frac{I_n+\alpha \breve C}{2N} \;,
\end{equation}
where
\begin{equation}
\label{E:Cdef}
    \breve C \equiv
    \begin{pmatrix}
        0 & \breve U_n^\dag \\
        \breve U_n & 0
    \end{pmatrix} \;.
\end{equation}
Using the binomial theorem, we can expand
$\tr(\breve\rho^{\,s}_\alpha)$ in terms of $\tr(\breve C^k)$:
\begin{equation}
\label{E:binomialform}
    \tr(\breve\rho^{\,s}_\alpha) =
    \left({1\over2N}\right)^{\! \! s} \sum_{k=0}^s {s \choose k} \alpha^k \tr(\breve C^k) \;.
\end{equation}
When $k$ is odd, $\breve C^k$ is block off-diagonal, so its trace
vanishes.  When $k$ is even, we have
\begin{equation}
\tr(\breve C^k) = 2\,\tr\Bigl((\breve U_n \breve U_n^\dag)^{k/2}\Bigr)\;.
\end{equation}
When $k=2$, this simplifies to $\tr(\breve C^2)=2\tr(\breve U_n\breve
U_n^\dag)=2\tr(U_n U_n^\dag)=2N$.  The crucial step here follows
immediately from the property $\tr(\breve A\breve B)=\tr(AB)$, which we
prove as a Lemma in the Appendix.  Note that in general $\tr(\breve A_1
\breve A_2 \ldots \breve A_l) \not= \tr(A_1 A_2 \ldots A_l)$ if $l>2$,
so we cannot give a similar general calculation of
$\tr(\breve\rho^{\,s}_\alpha)$ for even $s \ge 4$, since it involves terms
of this form.

Using Eq.~(\ref{E:binomialform}), we can now obtain three independent
constraint equations on the eigenvalues
$\lambda_j=\lambda_j(\breve\rho_\alpha)$ of the partial transpose
$\breve\rho_\alpha$:
\begin{equation} \label{E:sumsofpowers}
    \sum_{j=1}^{2N}\lambda_j^s=
    \tr(\breve \rho^{\,s}_\alpha)
    = \frac{1}{2^s N^{s-1}}[(1+\alpha)^s + (1-\alpha)^s]\;,\qquad s=1,2,3.
\end{equation}
Since the negativity is given by
\begin{equation} \label{E:negdef}
    \sM(\rho_\alpha) = \sum_j \abs{\lambda_i} \;,
\end{equation}
we can find an upper bound on the negativity by maximizing
$\sum_j\abs{\lambda_j}$ subject to the
constraints~(\ref{E:sumsofpowers}).  If we consider only the $s=1,2$
constraints, we obtain a nontrivial upper bound on the negativity with
little effort.  We find that adding the constraint $s=3$ adds nothing
asymptotically for large $N$, but for small $N$ yields a tighter bound
than we get from the $s=1,2$ constraints, although this comes at the
cost of considerably more effort.  We emphasize that these bounds apply
to all bipartite divisions and to all unitaries $U_n$.  Notice that we
have no reason to expect these bounds to be saturated, since the traces
of higher powers of $\breve\rho_\alpha$ impose additional constraints
that we are ignoring.  The one exception is the case of three qubits,
where the $s=1,2,3$ constraints are a complete set, and indeed, in this
case, the $s=1,2,3$ bound is $5/4$, which is saturated by the unitary
found in Sec.~\ref{S:examples}.

\begin{figure}[t]
\begin{center}
\includegraphics[scale=.6]{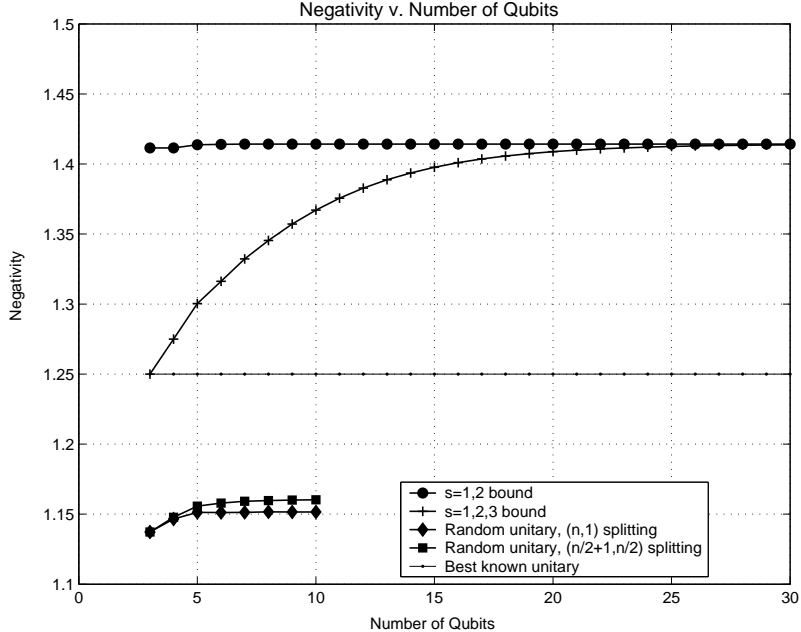}
\caption{Plot of the bounds on the negativity of states of the
form~(\ref{E:rhoout}), i.e., for a pure-state input in the zeroth
register ($\alpha=1$).  The uppermost plot is the simple analytic bound
$\sM_{1,2}=\sqrt2$, obtained using the $s=1,2$~constraint equations;
the next largest plot is the numerically constructed $s=1,2,3$ bound.
One can see that the $s=1,2,3$ bound asymptotes to the $s=1,2$ bound.
As noted in the text, these bounds are independent of the unitary $U_n$
and the bipartite division.  The flat line shows the negativity $5/4$
for the state constructed in Sec.~\ref{S:examples}, currently the state
of the form~(\ref{E:rhoout}) with the largest demonstrated negativity;
notice that for $n+1=3$, this state attains the $s=1,2,3$ bound.  The
lowest two sets of data points display the expected negativities for a
randomly chosen unitary using the bipartite splittings $(n,1)$ and
$\left(\left\lfloor n/2 \right\rfloor +1, \left\lceil n/2
\right\rceil\right)$, which were also plotted in Fig.~\ref{F:random}.}
\label{F:totalfig}
\end{center}
\end{figure}

The remainder of this section is devoted to calculating the $s=1,2$ and
$s=1,2,3$ upper bounds.  A graphical summary of our results is for
$\alpha=1$ presented in Fig.~\ref{F:totalfig}.

\subsection{The $s=1,2$ Bound}

We can use Lagrange multipliers to reduce the problem to maximizing a
function of one variable, but first we must deal with the absolute
value in Eq.~(\ref{E:negdef}).  To do so, we assume that $t$ of the
eigenvalues are negative and the $2N-t$ others are nonnegative, where
$t$ becomes a parameter that must now be included in the maximization.
We want to maximize
\begin{equation}
\sM_{1,2} = -\sum_{i=1}^t \lambda_i + \sum_{j=t+1}^{2N} \lambda_j\;,
\end{equation}
 subject to the constraints
\begin{equation} \label{E:cons}
    \sum_{k=1}^{2N} \lambda_k = 1
    \quad \mbox{and} \quad \sum_{k=1}^{2N}
    \lambda_k^2 = \frac{1+\alpha^2}{2N} \;.
\end{equation}
The notation we adopt here for the indices is that $i$ labels negative
eigenvalues and $j$ labels nonnegative eigenvalues, while $k$ can label
either.  This serves to remind us of the sign of an eigenvalue just by
looking at its index.

Introducing Lagrange multipliers $\mu$ and $\nu$, the function we want
to maximize is
\begin{equation}
    f(\lambda_k,t) =
    -\sum_{i=1}^t \lambda_i + \sum_{j=t+1}^{2N} \lambda_j +
    \mu\!\left(\sum_{k=1}^{2N} \lambda_k -1\right)
    + \nu\!\left(\sum_{k=1}^{2N} \lambda_k^2 -\frac{1+\alpha^2}{2N}\right)\;.
\end{equation}
Differentiating with respect to $\lambda_i$ and then $\lambda_j$, we find
\begin{eqnarray}
    -1+\mu+2\nu \lambda_i = 0 \;, \\
    +1+\mu+2\nu \lambda_j = 0 \;.
\end{eqnarray}
We immediately see that in the maximal solution, all the negative
eigenvalues are equal, and all the nonnegative eigenvalues are equal.
We can now reformulate the problem in the following way.  If we call
the two eigenvalues $\lambda_-$ and $\lambda_+$, our new problem is to
maximize
\begin{equation} \label{E:newneg}
    \sM_{1,2} = \sum_k \abs{\lambda_k}= -t\lambda_-+(2N-t)\lambda_+\;,
\end{equation}
subject to the constraints
\begin{eqnarray} \label{E:constraint1}
    t \lambda_- + (2N-t) \lambda_+ &=& 1\;,\\
    t \lambda_-^2 + (2N-t) \lambda_+^2 &=& \frac{1+\alpha^2}{2N}\;.
    \label{E:constraint2}
\end{eqnarray}
We can now do the problem by solving the constraints for $\lambda_-$
and $\lambda_+$ in terms of $t$, plugging these results into
$\sM_{1,2}$, and then maximizing over $t$.

Before continuing, we note two things.  First, $t$ cannot be $2N$, for
if it were, then all the eigenvalues would be negative, making it
impossible to satisfy Eq.~(\ref{E:constraint1}).  Second, unless
$\alpha=0$, $t$ cannot be 0, for if it were, then all the eigenvalues
would be equal to $1/2N$ by Eq.~(\ref{E:constraint1}), a situation
Eq.~(\ref{E:constraint2}) says can occur only if $\alpha=0$.  Since we
are not really interested in the case $\alpha=0$, for which
$\rho_\alpha$ is always the maximally mixed state, we assume $\alpha>0$
and $0<t<2N$ in what follows.

Solving Eqs.~(\ref{E:constraint1}) and (\ref{E:constraint2}) and
plugging the solutions into Eq.~(\ref{E:newneg}), we get the two
solutions
\begin{equation}
\label{E:ellipse}
    \sM_{1,2} = \frac{N-t \pm \alpha \sqrt{t(2N-t)}}{N} \;.
\end{equation}
We choose the positive branch, since it contains the maximum.
Maximizing with respect to $t$ treated as a continuous variable, we
obtain the upper bound,
\begin{equation} \label{E:maxneg}
    \sM_{1,2} = \sqrt{1+\alpha^2}\, \stackrel{\alpha \to 1}{=}
    \sqrt 2 \simeq 1.414\;,
\end{equation}
which occurs when the degeneracy parameter is given by
\begin{equation}
\label{E:degeneracy}
    t = N\!\left(1 -\frac{1}{\sqrt{1+\alpha^2}} \right)
    \stackrel{\alpha \to 1}{\simeq} 0.292\,N \;.
\end{equation}
The numbers on the right are for the case $\alpha = 1$, corresponding
to the special qubit starting in a pure state.  Notice that the upper
bound~(\ref{E:maxneg}) allows a negativity greater than 1 for all
$\alpha$ except $\alpha=0$.

Since we did not yet enforce the condition that $t$ be a positive
integer, the bound~(\ref{E:maxneg}) can be made tighter for specific
values of $N$ and $\alpha$ by calculating $t$ and checking which of the
two nearest integers yields a larger $\sM_{1,2}$.  Asymptotically,
however, the ratio $t/N$ can approach any real number, so this bound
for continuous $t$ is the same as the bound for integer $t$ in the
limit $N\to\infty$.

\subsection{The $s=1,2,3$ Bound}

To deal with this case, we again make the assumption that $t$ of the
eigenvalues are negative and $2N-t$ are nonnegative and thus write
\begin{equation}
\sM_{1,2,3} = -\sum_{i=1}^t \lambda_i + \sum_{j=t+1}^{2N} \lambda_j\;,
\end{equation}
as before.  In addition to the constraints~(\ref{E:cons}), we now have a
third constraint
\begin{equation}
\label{E:constraint3}
    \sum_{k=1}^{2N} \lambda_k^3 = \frac{1+3 \alpha^2}{4N^2} \;.
\end{equation}
We specialize to the case $\alpha = 1$ for the remainder of this
subsection, because it is our main interest, and the algebra for
the general case becomes difficult.

Introducing three Lagrange multipliers, we can write the function we
want to maximize as
\begin{equation}
\label{E:lagrange123}
    f(\lambda_k,t) =
    -\sum_{i=1}^t \lambda_i + \sum_{j=t+1}^{2N} \lambda_j +
    \mu\!\left( \sum_{k=1}^{2N} \lambda_k - 1 \right)+
    \nu\!\left( \sum_{k=1}^{2N} \lambda_k^2 - \frac{1}{N} \right) +
    \xi\!\left( \sum_{k=1}^{2N} \lambda_k^3 - \frac{1}{N^2} \right)\;.
\end{equation}
Differentiating with respect to $\lambda_i$ and then $\lambda_j$ gives
\begin{eqnarray}
    -1+\mu+2\nu \lambda_i + 3 \xi \lambda_i^2 &=& 0 \;, \\
    +1+\mu+2\nu \lambda_j + 3 \xi \lambda_j^2 &=& 0 \;.
\end{eqnarray}
These equations being quadratic, we see that there are at most two
distinct negative eigenvalues and at most two distinct nonnegative
eigenvalues.  Since the sum of the two solutions of either of these
equations is $-2\nu/3\xi$, however, we can immediately conclude either
that one of the potentially nonnegative solutions is negative or that
one of the potentially negative solutions is positive.  Hence, we find
that at least one of the four putative eigenvalues has the wrong sign,
implying that there are at most three distinct eigenvalues, though we
don't know whether one or two of them are negative.

Labelling the three eigenvalues by $A$, $B$, and $C$, we can reduce the
problem to solving the three constraint equations,
\begin{eqnarray} \label{E:123}
    u A + v B + w C & = & 1\;,\nonumber \\
    u A^2 + v B^2 + w C^2 & = & 1/N\;,\\
    u A^3 + v B^3 + w C^3 & = & 1/N^2\;,\nonumber
\end{eqnarray}
for $A$, $B$, and $C$ and then maximizing $\sM_{1,2,3}$ over the
degeneracy parameters $u$, $v$, and $w$, which are nonnegative positive
integers satisfying the further constraint
\begin{equation}
u+v+w=2N\;.
\end{equation}
We do not associate any particular sign with $A$, $B$, and $C$; the
signs are determined by the solution of the equations.

One might hope that the symmetry of Eqs.~(\ref{E:123}) would allow for
a simple analytic solution, but this appears not to be the case. In
solving the three equations, one is inevitably led to a sixth-order
polynomial in one of the variables, with the coefficients given as
functions of $u$, $v$, and $w$.  Rather than try to solve this
equation, which appears intractable, we elected to do a brute force
optimization for any given value of $2N$ by solving Eqs.~(\ref{E:123})
for each possible value of $u$, $v$, and $w$.  Picking the solution
that has the largest negativity then yields the global maximum.  We did
this for each $N$ up to $2N=78$.  The values of
$u$, $v$, and $w$ that maximize the negativity are always
\begin{equation} \label{E:uvw}
    u=\left[N\left(1-\frac{1}{\sqrt 2}\right)\right] \;,
    \ v=1 \;,
    \ w = 2N-1-u \;,
\end{equation}
where $[x]$ denotes the integer nearest to $x$.  The unique eigenvalue
corresponding to $v=1$ is the largest positive eigenvalue, $w$ is the
degeneracy of another positive eigenvalue, and $u$ is the degeneracy of
the negative eigenvalue.  Notice that the degeneracy of the negative
eigenvalue is exactly what was found in the $s=1,2$ case.  Using the
results~(\ref{E:uvw}) as a guide, we did a further numerical
calculation of the maximum for larger values of $N$, by considering
only the area around the degeneracy values given by
Eq.~(\protect\ref{E:uvw}).  While this is not a certifiable global
maximum, the perturbation expansion described below matches so well
that the two are indistinguishable if they are plotted together for
$n+1>7$.  This gives us confidence that the numerically determined
upper bound $\sM_{1,2,3}$, which we plot in Fig.~\ref{F:totalfig}, is
indeed a global maximum for all $N$.

We have used the numerical work to help formulate a perturbation
expansion that gives the first correction to the $N\to\infty$ behavior
of the $s=1,2,3$ bound.  Defining $x=1/N$, we rewrite the constraint
equations~(\ref{E:123}) as
\begin{eqnarray} \label{E:con}
    a A + b B + c\,C & = & x\;,\nonumber \\
    a A^2 + b B^2 + c\,C^2 & = & x^2\;,\\
    a A^3 + b B^3 + c\,C^3 & = & x^3\;,\nonumber
\end{eqnarray}
where $a=u/N$, $b=v/N$, and $c=w/N$.  We also have the constraint
\begin{equation} \label{E:con2}
a+b+c=2\;.
\end{equation}
As $x$ is the variable that is asymptotically small, we seek an
expansion in terms of it.

Our numerical work tells us that there are two positive eigenvalues,
one of which is larger and nondegenerate.  In formulating our
perturbation expansion, we let $B$ and $C$ be the positive eigenvalues,
with $B$ being the larger one, having degeneracy $v=b_1\ge1$.  We do
not assume that $b_1$ is 1, as the numerics show, but rather let the
equations force us to that conclusion.  With this assumption, the form
of the constraints~(\ref{E:con}) shows that the variables have the
following expansions to first order beyond the $N\to\infty$ form:
\begin{eqnarray} \label{E:degs}
    a & = & a_0 + a_1 x^{1/3}\;,\nonumber \\
    b & = & b_1x\;,\\
    c & = & c_0 + c_1x^{1/3}\;,\nonumber
\end{eqnarray}
and
\begin{eqnarray} \label{E:evs}
    A & = & A_0 x + A_1 x^{4/3}\;,\nonumber \\
    B & = & B_1 x^{2/3}\;,\\
    C & = & C_0 x + C_1 x^{4/3}\;.\nonumber
\end{eqnarray}
We see that we are actually expanding in the quantity $y=x^{1/3}$.  In
terms of these variables, the negativity is given by
\begin{eqnarray}
 \sM_{1,2,3}&=&
 \frac{-a A + b B + c C}{x}\nonumber\\
 &=&-a_0A_0+c_0C_0+(-a_0A_1-a_1A_0+c_0C_1+c_1C_0)x^{1/3}+O(x^{2/3})\;,
\end{eqnarray}
which we now endeavor to maximize.

Substituting Eqs.~(\ref{E:degs}) and (\ref{E:evs}) into the
constraints~(\ref{E:con}) and (\ref{E:con2}) and equating terms with
equal exponents of $x$, we obtain, to zero order,
\begin{eqnarray}
    a_0 + c_0 & = & 2\;,\nonumber \\
    a_0 A_0 + c_0 C_0 & = & 1\;,\\
    a_0 A_0^2+ c_0 C_0^2 & = & 1 \;.\nonumber
\end{eqnarray}
Solving for $a_0$, $c_0$, and $C_0$ in terms of $A_0$ and substituting
the results into the zero-order piece of $\sM_{1,2,3}$ gives
\begin{equation}
 \label{E:neg1}
 \sM_{1,2,3}=\frac{1-4A_0+2A_0^2}{1-2A_0+2A_0^2}\;.
\end{equation}
Maximizing Eq.~(\ref{E:neg1}) gives $A_0^2=1/2$ and, hence,
$A_0=-1/\sqrt{2}$, since $A$ is the negative eigenvalue.  This leads to
$a_0=1-1/\sqrt{2}$, $c_0 = 1+1/\sqrt{2}$, and $C_0=1/\sqrt{2}$, and the
resulting $N\to\infty$ upper bound is $\sM_{1,2,3}=\sqrt2$, as
expected.

If we carry this process out to first order beyond the $N\to\infty$
behavior, we obtain, after some algebraic manipulation, $\sM_{1,2,3} =
\sqrt2-b_1^{1/3}x^{1/3}/2^{7/6}+O(x^{2/3})$.  Maximizing this simply
means making $b_1$ as small as possible, i.e., choosing $b_1=1$, whence
we obtain the following asymptotic expression for the $s=1,2,3$ upper
bound:
\begin{equation}
\sM_{1,2,3} = \sqrt2 - \frac{1}{2^{7/6}N^{1/3}} +
O\!\left(\frac{1}{N^{2/3}}\right)\;.
\end{equation}
This shows that the upper bound of $\sqrt 2$ is approached
monotonically from below in the asymptotic regime.  In addition, the
procedure verifies that in the maximum solution, the largest positive
eigenvalue is nondegenerate.  For the case of qubits we have $N=2^n$,
implying that the approach to the $N\to\infty$ bound is exponentially
fast.

\section{Conclusion} \label{S:conclusion}

The mixed-state quantum circuit~(\ref{E:circuitalpha}) provides an
efficient method for estimating the normalized trace of a unitary
operator, a task that is thought to be exponentially hard on a
classical computer.  If one believes that global entanglement is the
essential resource for the exponential speedup achieved by quantum
computation, then the question begging to be answered is whether there
is any entanglement in the circuit's output
state~(\ref{E:rhooutalpha}). The purpose of this paper was to
investigate this question.

A notable feature of the circuit~(\ref{E:circuitalpha}) is that it
provides an efficient method for estimating the normalized trace no
matter how small the initial polarization $\alpha$ of the special qubit
in the zeroth register, as long as that polarization is not zero.
Since all the other qubits are initially completely unpolarized, we are
led to characterize the computational power of this circuit as the
``power of even the tiniest fraction of a qubit.''  We provide
preliminary results regarding the entanglement that can be achieved for
$\alpha<1$. Our results are consistent with, but certainly do not
demonstrate the conclusion that separable states cannot provide an
exponential speedup and that entanglement is possible no matter how
small $\alpha$ is.  The question of entanglement for subunity
polarization of the special qubit deserves further investigation.

Our key conclusions concern the case where the special qubit is
initially pure ($\alpha=1$).  We find that the
circuit~(\ref{E:circuit}) typically does produce global entanglement,
but the amount of this entanglement is quite small.  Using
multiplicative negativity to measure the amount of entanglement, we
show that as the number of qubits becomes large, the multiplicative
negativity in the state~(\ref{E:rhoout}) is a vanishingly small
fraction of the maximum possible multiplicative negativity for roughly
equal splittings of the qubits.  This hints that the key to
computational speedup might be the global character of the
entanglement, rather than the amount of the entanglement.  In the
spirit of the pioneering contribution of Wyler~\cite{Wyler74}, what
happier motto can we find for this state of affairs than {\it Multum ex
Parvo}, or A Lot out of A Little.

\begin{acknowledgments}
The authors thank H.~N. Barnum, A.~Denney, B.~Eastin, K.~Manne, N.~C.
Menicucci, and A.~Silberfarb for useful
discussions.  The Matlab code used to calculate the results of
Sec.~\ref{S:random} made use of T.~S.~Cubitt's freely available
algorithm for taking the partial transpose of a matrix; this and other
useful algorithms written by Cubitt are available at the website
\href{http://www.dr-qubit.org/matlab.html}{http://www.dr-qubit.org/matlab.html}.
The quantum circuits in this paper were typeset using Qcircuit, which
is freely available online at
\href{http://info.phys.unm.edu/Qcircuit}{http://info.phys.unm.edu/Qcircuit}.
This work was supported in part by US Army Research Office Contract
No.~W911NF-04-1-0242.
\end{acknowledgments}

\appendix
\section{Proof of the Lemma}{\label{A:appendix}}

\textbf{Lemma:} $\tr(\breve A \breve B) = \tr(A B) $.\\

\textbf{Proof:} Define two operators $A$ and $B$ by
\begin{equation}
A=\sum_{i,j,k,l} a_{ij,kl} \ket{ij}\!\bra{kl}\;,
\quad
B=\sum_{m,n,p,q} b_{mn,pq} \ket{mn}\!\bra{pq} \;.
\end{equation}
Taking the partial transpose with respect to the second subsystem, we
find
\begin{equation}
\breve A=\sum_{i,j,k,l} a_{ij,kl} \ket{il}\!\bra{kj}\;,
\quad
\breve B=\sum_{m,n,p,q} b_{mn,pq} \ket{mq}\!\bra{pn} \;.
\end{equation}
Calculating the quantities of interest, we find that they are indeed equal.
\begin{eqnarray}
    \tr(\breve A \breve B) &=&
        \sum_{\scriptsize\begin{array}{l}i,j,k,l, \\ m,n,p,q \end{array}}
        a_{ij,kl} b_{mn,pq} \braket{pn}{il} \braket{kj}{mq}
    = \sum_{i,j,k,l} a_{ij,kl} b_{kl,ij} \;, \\
    \tr\left(A  B\right) &=&
        \sum_{\scriptsize \begin{array}{l}i,j,k,l, \\ m,n,p,q \end{array}}
        a_{ij,kl} b_{mn,pq} \braket{pq}{ij} \braket{kl}{mn}
    = \sum_{i,j,k,l} a_{ij,kl} b_{kl,ij}\;.
\end{eqnarray}


\begin{references}

\bibitem{DiVincenzo00} D.~DiVincenzo, Fortschr.\ Phys. {\bf 48}, 9 (2000).

\bibitem{Jones01} J.~A. Jones, Prog. NMR Spectroscopy {\bf 38}, 325 (2001).

\bibitem{Schulman99} L.~J. Schulman and U.~V. Vazirani, in Proceedings
of the 31st Annual ACM Symposium on Theory of Computing (ACM Press, New
York, 1999), p.~322 (extended version available as {\tt arXiv.org
e-print \href{http://arxiv.org/abs/quant-ph/9804060}{quant-ph/9804060}}).

\bibitem{kl98} E.~Knill and R.~Laflamme, Phys.\ Rev.\ Lett. {\bf 81}, 5672 (1998).

\bibitem{Ambainis00} A.~Ambainis, L.~J. Schulman, and U.~V. Vazirani,
in Proceedings of the 32nd Annual ACM Symposium on Theory of Computing
(ACM Press, New York, 2000), p.~697.

\bibitem{Laflamme02} R.~Laflamme, D.~G. Cory, C.~Negrevergne, and L.~Viola,
Quant.\ Inf.\ Comp.\ {\bf 2}, 166 (2002).

\bibitem{pklo04} D.~Poulin, R.~Blume-Kohout, R.~Laflamme, and H.~Ollivier,
Phys.\ Rev.\ Lett.\ {\bf 92}, 177906 (2004).

\bibitem{elpc04} J.~Emerson, S.~Lloyd, D.~Poulin, and D.~Cory, Phys.\ Rev.~A
{\bf 69}, 050305(R) (2004).

\bibitem{Josza99} R.~Josza and N.~Linden, Proc.~Roy.~Soc.~Lond.~A {\bf 459},
2011 (2003).

\bibitem{N&C} M.~Nielsen and I.~Chuang, {\it Quantum Computation and Quantum
Information\/} (Cambridge University Press, Cambridge, England, 2000).

\bibitem{b99} S.~L. Braunstein, C.~M. Caves, R.~Jozsa, N.~Linden, S.~Popescu,
and R.~Schack, Phys.\ Rev.\ Lett.\ {\bf 83}, 1054 (1999).

\bibitem{gb03} L.~Gurvits and H.~Barnum, Phys.\ Rev.~A {\bf 68},
042312 (2003).

\bibitem{gb04} L.~Gurvits and H.~Barnum, {\tt arXiv.org e-print
\href{http://arxiv.org/abs/quant-ph/0409095}{quant-ph/0409095}}.

\bibitem{dur99} W.~D{\"u}r, J.~I. Cirac, and R.~Tarrach, Phys.\ Rev.\ Lett.\
{\bf 83}, 3562 (1999).

\bibitem{p96} A.~Peres,  Phys.\ Rev.\ Lett.\ {\bf 77}, 1413 (1996).

\bibitem{hhh96} M.~Horodecki, P.~Horodecki, R.~Horodecki, Phys.~Lett.~A
{\bf 223}, 1 (1996).

\bibitem{Vidal02} G.~Vidal and R.~F. Werner, Phys.\ Rev.~A {\bf 65}, 032314
(2002).

\bibitem{dhhmno05} C.~M.~Dawson, H.~L.~Haselgrove, A.~P.~Hines, D.~Mortimer,
M.~A.~Nielsen, and T.~J.~Osborne, Quant.\ Inf.\ Comp.\ {\bf 5}, 102 (2005).

\bibitem{ek90} A.~Ehrenfeucht and  M.~Karpinski, Technical Report~tr-90-033,
International Computer Science Institute, Berkeley, CA (1990),
available online at {\tt\href{http://www.citeseer.com}{http://www.citeseer.com}}.

\bibitem{Plenio05} M.~B.~Plenio, {\tt arXiv.org e-print
\href{http://arxiv.org/abs/quant-ph/0505071}{quant-ph/0505071}}.

\bibitem{SLee03}
S.~Lee, D.~P.~Chi, S.~D.~Oh, and J.~Kim, Phys.\ Rev.~A {\bf 68}, 062304 (2003).

\bibitem{Conway90} J.~Conway, {\it A Course in Functional Analysis}
(New York, Springer-Verlag, 1990).

\bibitem{Hurwitz1897} A.~Hurwitz, Nachr.\ Ges.\ Wiss.\ G\"ott.\ Math.-Phys.~{\bf Kl},
71 (1897).  The Hurwitz decomposition is discussed in the appendix of
M.~Po\'zniak, K.~\.Zyczkowski, and M.~Ku\'s, J.~Phys.~A:
Math.~Gen.~{\bf 31}, 1059 (1998).

\bibitem{Emerson03} J.~Emerson, Y.~S.~Weinstein, M.~Saraceno, S.~Lloyd,
and D.~G.~Cory, Science {\bf 302}, 2098 (2003).

\bibitem{Scott03} A.~J.~Scott and C.~M.~Caves, J.~Phys.~A {\bf 36}, 9553
(2003).

\bibitem{Wyler74} J.~A.~Wyler, Gen.~Rel.~Grav. {\bf 5}, 175 (1974).


\end{references}
\end{document}